\documentclass[twocolumn]{aastex631}
\usepackage{gensymb}
\usepackage{amsmath}

\shorttitle{Superhabitability of High-Obliquity and High-Eccentricity Planets}
\shortauthors{Jernigan et al.}

\graphicspath{{./}{figures/}}

\begin{document}

\title{Superhabitability of High-Obliquity and High-Eccentricity Planets}
\author[0000-0003-0410-0909]{Jonathan Jernigan}
\affiliation{Department of Earth, Atmospheric, and Planetary Science, Purdue University, West Lafayette, IN 47901, USA}
\author[0000-0002-8041-3184]{Émilie Laflèche}
\affiliation{Department of Earth, Atmospheric, and Planetary Science, Purdue University, West Lafayette, IN 47901, USA}
\author[0000-0001-6454-1883]{Angela Burke}
\affiliation{Department of Earth, Atmospheric, and Planetary Science, Purdue University, West Lafayette, IN 47901, USA}
\author[0000-0002-3249-6739]{Stephanie Olson}
\affiliation{Department of Earth, Atmospheric, and Planetary Science, Purdue University, West Lafayette, IN 47901, USA}

\correspondingauthor{Stephanie Olson}
\email{stephanieolson@purdue.edu}

\begin{abstract}
Planetary obliquity and eccentricity influence climate by shaping the spatial and temporal patterns of stellar energy incident at a planet's surface, affecting both the annual mean climate and magnitude of seasonal variability. Previous work has demonstrated the importance of both planetary obliquity and eccentricity for climate and habitability, but most studies have not explicitly modeled the response of life to these parameters. While exaggerated seasons may be stressful to some types of life, a recent study found an increase in marine biological activity for moderately high obliquities \textless 45$^{\circ}$ assuming an Earth-like eccentricity. However, it is unclear how life might respond to obliquities \textgreater 45$^{\circ}$, eccentricities much larger than Earth's, or the combination of both. To address this gap, we use cGENIE-PlaSim, a 3-D marine biogeochemical model coupled to an atmospheric general circulation model, to investigate the response of Earth-like marine life to a large range of obliquities (0--90$^{\circ}$) and eccentricities (0--0.4). We find that marine biological activity increases with both increasing obliquity and eccentricity across the parameter space we considered, including the combination of high obliquity and high eccentricity. We discuss these results in the context of remote biosignatures, and we argue that planets with high obliquity and/or eccentricity may be superhabitable worlds that are particularly favorable for exoplanet life detection.
\end{abstract}

\keywords{Exoplanets (498) --- Habitable planets (695) ---  Planetary climates (2184) --- Astrobiology (74) --- Biosignatures (2018)}

\section{Introduction}

\begin{figure*}[ht!]
\gridline{\fig{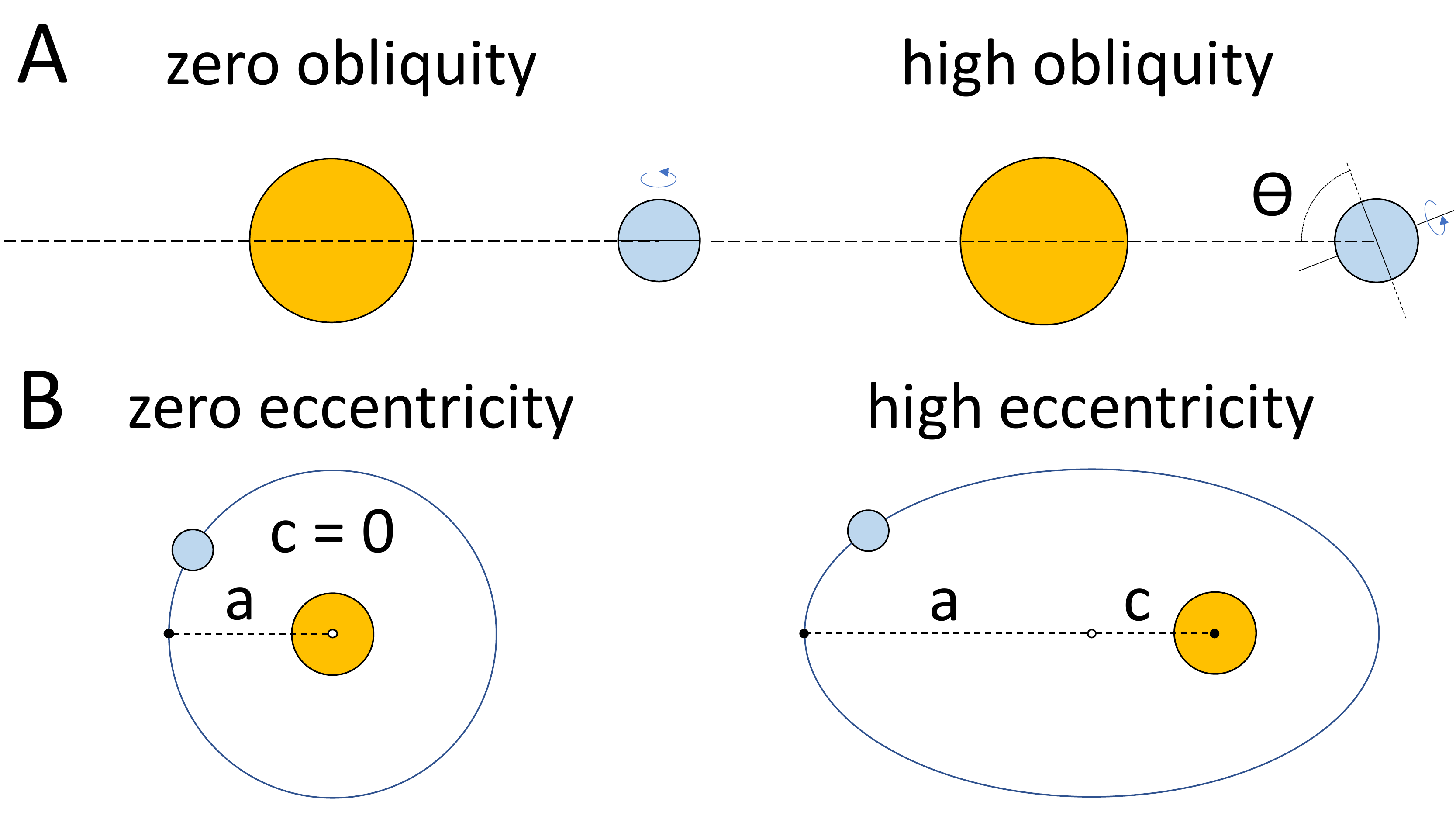}{0.5\textwidth}{}\fig{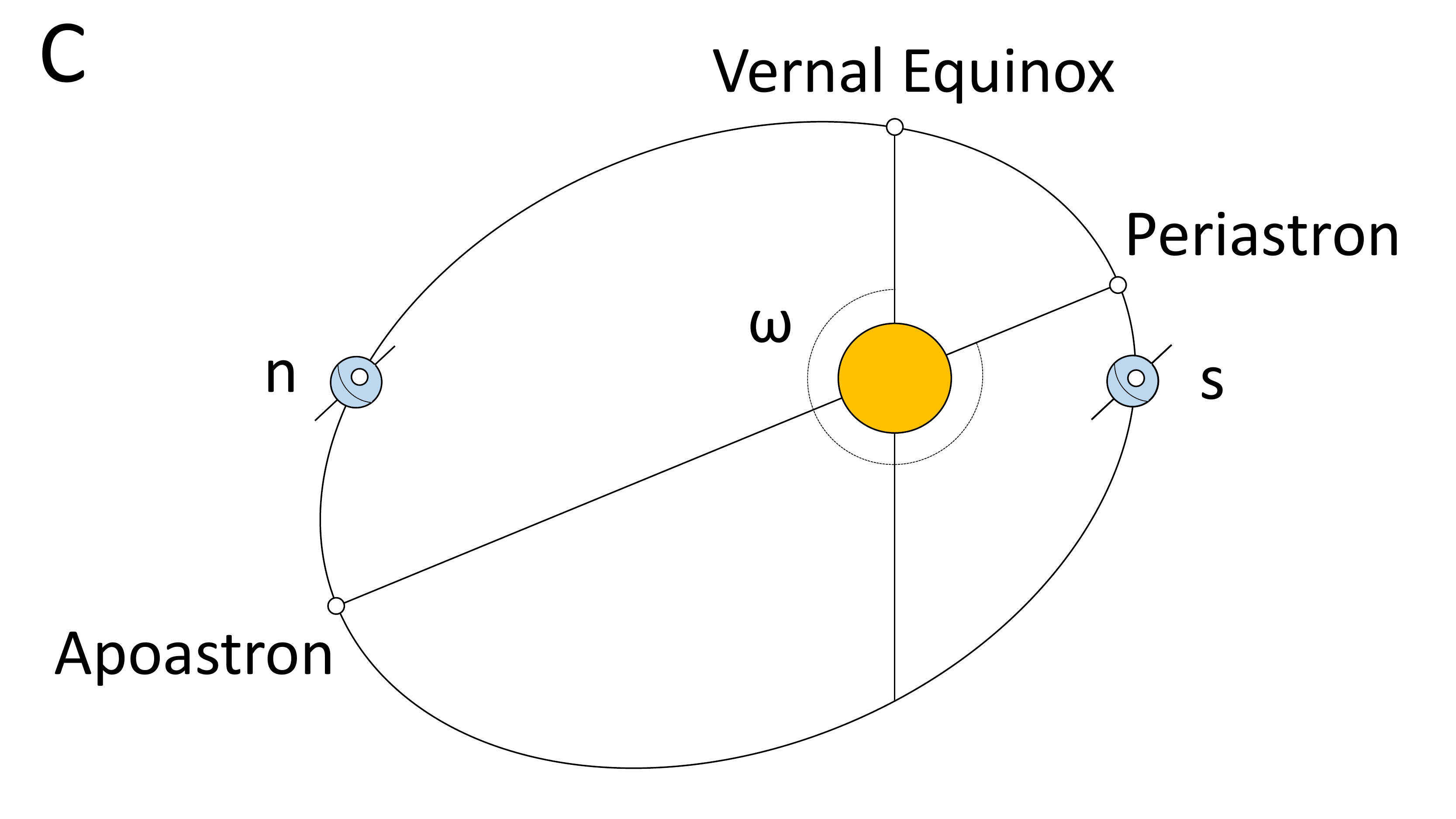}{0.5\textwidth}{}}
\caption{A: Planetary obliquity ($\theta$) is the angle between a planet's orbital and equatorial planes. B: The eccentricity of an elliptical orbit is the ratio of the distance between one focus and the geometric center of the ellipse (c) to the semi-major axis (a). C: The argument of the periastron ($\omega$) is the angle between the vernal equinox and the periastron. For $\omega = 282.7^{\circ}$, the present-day Earth value, the Northern Hemisphere (NH) summer occurs near the apoastron, while the Southern Hemisphere (SH) summer occurs near the periastron (Table \ref{tab:Orbit}).}
\label{fig:orbits}
\end{figure*}

Planetary obliquity and orbital eccentricity modulate planetary climate, in part by generating seasons due to time-varying instellation patterns. The effects of these orbital parameters on climate are modest for Earth, but a diversity of annual mean climate and seasonality scenarios are possible on other worlds.
The diversity of planetary obliquities is apparent even among terrestrial planets in our own solar system, with Venus and Mercury exhibiting obliquities of 180$^\circ$ and near 0$^\circ$ respectively, while Mars' obliquity varies chaotically from 0--60$^\circ$ \citep{laskar_chaotic_1993}. 
A similarly large range of obliquities is expected in other stellar system architectures as a result of complex orbital interactions \citep{millholland_obliquity-driven_2019} and giant impacts \citep{li_planetary_2020}. 
Eccentricity is less varied among observed low-mass planets to date, including terrestrial planets in the Habitable Zone (HZ) \citep{van_eylen_eccentricity_2015, eylen_orbital_2019, guendelman_atmospheric_2020}. Nonetheless, eccentricity varies among solar system planets and known exoplanets \citep{udry_statistical_2007}, and these differences in eccentricity are expected to impact planetary climate, seasonal cycles, and habitability \citep{williams_earth-like_2002, dressing_habitable_2010}.
Considering the diversity of obliquities and eccentricities anticipated among HZ planets, many potentially habitable exoplanets may experience amplified seasonality compared to Earth. 
This enhanced seasonality may have broad-reaching consequences for planetary habitability and exoplanet life detection efforts.

Exaggerated seasonality on high-obliquity and/or high-eccentricity worlds could be challenging for life. However, higher obliquity planets have warmer poles and more equable climates than lower obliquity planets due to decreased planetary albedo from a reduction in both ice and cloud cover, which may allow higher obliquity planets to maintain surface liquid water at farther distances from their host star \citep{williams_habitable_1997, williams_extraordinary_2003, spiegel_habitable_2009, dressing_habitable_2010, armstrong_effects_2014, ferreira_climate_2014, linsenmeier_climate_2015, wang_effects_2016, kilic_stable_2018, nowajewski_atmospheric_2018, guendelman_atmospheric_2019, kang_mechanisms_2019, colose_enhanced_2019, palubski_habitability_2020, komacek_constraining_2021}. Some authors have even suggested that planets with higher obliquities may be `superhabitable' worlds more suitable for life than Earth \citep{heller_superhabitable_2014, olson_oceanographic_2020}, but most models for planetary habitability to date lack explicit representation of life.

Recent biogeochemical modeling suggests that moderately high planetary obliquity enhances the productivity of marine life \citep{barnett_moderately_2022}. However, \citet{barnett_moderately_2022} only considered obliquities up to 45$^\circ$. It is thus unknown whether marine life would respond positively to even higher obliquities, or if there exists an optimal obliquity beyond which strong seasonal contrasts begin to negatively affect marine life. Their study also neglects the effects of seasonality arising from eccentricity. 

The mechanism by which obliquity influenced biospheric productivity in \citet{barnett_moderately_2022}'s study was through enhanced nutrient recycling due to seasonal breakdown of the ocean's thermal stratification, which otherwise tends to limit vertical mixing. This same phenomenon may occur on planets with eccentric orbits, but asymmetry in season duration may yield different dynamical and biological effects. Planets on highly eccentric orbits can also move into and out of their star's HZ with each orbit bearing unclear consequences for habitability and life. Ultimately, it is unknown how marine life might respond to the large range of orbital scenarios we expect to find with current and next-generation observatories. 

We address this gap here by using an atmospheric GCM (PlaSim) coupled to a 3D biogeochemical model (cGENIE) to simulate the climates of high-obliquity and high-eccentricity worlds and characterize the response of an Earth-like biosphere to each orbital scenario. 
We review how obliquity and eccentricity drive seasonality and provide an introduction to key biogeochemical processes regulating the biospheric response to seasons in Section \ref{background}. We then describe our climate and biogeochemical modeling approach in Section \ref{methods}. We present our results regarding the response of surface environments and life to seasons in Section \ref{results}, and we discuss implications for exoplanet habitability and life detection on high-obliquity and high-eccentricity worlds in Section \ref{discussion}. Finally, we summarize our findings and offer recommendations for future work in Section \ref{conclusions}.

\section{Background}\label{background}

\subsection{Effects of Orbital Parameters on Planetary Climate and Seasonality}
Planetary obliquity ($\theta$ in Figure \ref{fig:orbits}A) shapes the spatial and temporal distribution of incident stellar radiation, generating the familiar hemispheric seasons that we experience on Earth. At low obliquity (herein, 0--23.5$^\circ$), the equator receives the most stellar energy, while the poles receive relatively little. Seasonality is also limited, particularly at low latitudes. At moderate obliquity (herein, 23.5--54$^\circ$), the spatial distribution of stellar energy becomes more uniform on annual average, but with increasing temporal variability. The result is a climate with smaller equator-to-pole contrasts, less ice on annual average, and larger seasonal contrasts at all latitudes---despite receiving the same stellar energy flux on global average. At high obliquity (herein, 54--90$^{\circ}$), the poles receive more stellar energy than the equator, despite increasingly dramatic temporal variability in irradiance \citep{ward_climatic_1974}. In this scenario, equatorial ice belts (rather than polar ice caps) may become stable \citep{rose_ice_2017, kilic_stable_2018}. 

Eccentric orbits (Figure \ref{fig:orbits}B) generate seasons as the star-planet separation varies throughout the orbit. In this case, summer occurs when the planet is closer to its host star, while winter occurs while it is further away. Unlike seasons arising from obliquity, both hemispheres will experience the same season as the stellar energy received by the planet changes across the year. Moreover, the summer vs. winter seasons will differ in duration. This contrast arises due to evolving gravitational interaction between the planet and star as the orbital distance changes, as described by Kepler's second law. Consequently, the summer will be relatively short as the planet speeds up close to its star and the winter will be relatively long as the planet orbits slower further from its star. Both the magnitude of seasonal flux variations and the asymmetry in season duration increase with increasing eccentricity. 

For planets with a combination of both non-zero planetary obliquity and orbital eccentricity, both seasonality effects will manifest. These effects may be additive or opposing, depending on the the timing of obliquity-driven seasons relative to the point in the planet's eccentric orbit closest to the star (periastron) and the point furthest from the star (apoastron). This relationship can be described by the argument of the periastron ($\omega$ in Figure \ref{fig:orbits}C), i.e, the angle between the orbital position where both hemispheres receive identical instellation during the NH vernal equinox and the periastron. 

For $\omega =$ 0$^\circ$ or 180$^\circ$, the Northern Hemisphere (NH) vernal or autumnal equinox respectively will occur at the periastron, with the opposite occurring at apoastron. The effects of eccentricity and obliquity on the seasons will be out-of-phase from one another, with the global summer and winter generated by eccentricity corresponding with spring and autumn generated by obliquity. The resulting seasonality will therefore be subdued.

\begin{figure*}[ht]
\gridline{\fig{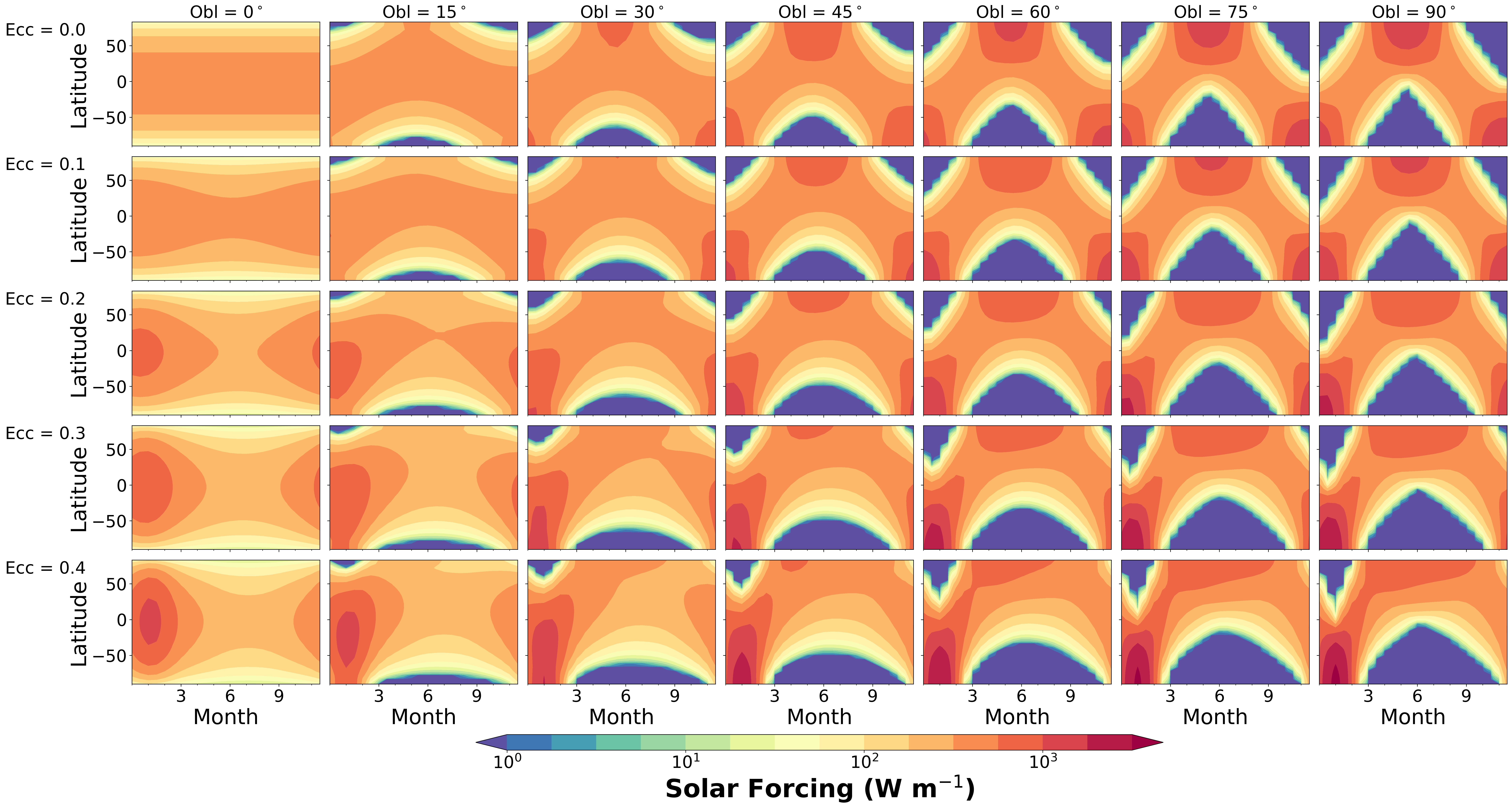}{\textwidth}{}}
\caption{Zonally averaged solar short wave forcing at top of atmosphere across the seasonal cycle for varied obliquity (columns) and eccentricity (rows). Time advances along the x-axis and latitude is represented on the y-axis for each subpanel. Solar forcing peaks at periastron and reaches a minimum at apoastron, the timing of which varies with eccentricity.}
\label{fig:solar}
\end{figure*}

For $\omega =$ 90$^\circ$ or 270$^\circ$, the NH summer solstice (SS) or winter solstice (WS) respectively occurs at the periastron, with the opposite occurring at the apoastron (Table \ref{tab:Orbit}). The global summers and winters generated by eccentricity and the hemispheric summers and winters generated by obliquity will coincide for one hemisphere and conflict for the other. The resulting seasonality will therefore be amplified for one hemisphere and muted for the other (Figure \ref{fig:solar}).

\begin{deluxetable}{ccccc}
\tablenum{1}
\tablecaption{Timing of Orbital Events (Month, Day)}
\tablewidth{\textwidth}
\tablehead{
\colhead{Eccentricity} & \colhead{Periastron} & \colhead{Apoastron} & \colhead{SS (NH)} & \colhead{WS (NH)}
\label{tab:Orbit}
}
\startdata
0.0 & - & - & (6, 20) & (12, 20) \\
0.1 & (1, 13) & (7, 14) & (6, 28) & (1, 3) \\
0.2 & (1, 24) & (7, 24) & (7, 6) & (1, 16) \\
0.3 & (2, 4) & (8, 4) & (7, 11) & (1, 27) \\
0.4 & (2, 14) & (8, 13) & (7, 15) & (2, 8) \\
\enddata
\end{deluxetable}

\subsection{Characterizing biological activity}
Photosynthesis by marine microorganisms produces biomass via the following chemical reaction: \begin{equation} \label{eqn:photo}  CO_2 + H_2O \xrightarrow{hv} CH_2O + O_2 \end{equation} where CH$_2$O is geochemical shorthand for biomass that in reality includes a number of other macro- (e.g., N, P) and micro-nutrients (e.g., Fe, Mo). These nutrients, especially P, typically limit photosynthetic rates (`primary productivity') on global and long-term average on Earth \citep{tyrrell_relative_1999}, but any of these ingredients can be locally or seasonally limiting. For instance, light may limit productivity during winter darkness while nutrients may limit productivity during photon-replete summers. We expect similar spatial and temporal variability in primary productivity on other worlds experiencing seasons.

The majority of biomass produced by photosynthesis is broken down in the surface wind-mixed layer by respiration: \begin{equation} \label{eqn:resp} CH_2O + O_2 \rightarrow CO_2 + H_2O \end{equation} However, a small portion sinks from the mixed layer into the deeper layers of the ocean. This flux is termed biological `export production', or simply `export', and is measured in teramoles of particulate organic carbon per year (Tmol POC yr$^{-1}$). Export production is typically greatest when and where primary production is greatest, but export is also sensitive to temperature and thus it is not necessarily a fixed fraction of primary productivity in either space or time. 

The process by which export POC is transferred from the surface to depth is referred to as the `biological pump'. The net consequence of the biological pump is to concentrate bioessential nutrients at depth while depleting them at the surface. As a result, the biosphere depends on vertical mixing to replenish nutrients to the sunlit regions of the water column where photosynthesis is viable, also referred to as the photic zone. Greater nutrient recycling ultimately allows greater biological productivity, increasing the production (but not necessarily the survival) of biogenic gases like O$_2$ and CH$_4$ that may enter the atmosphere and serve as `biosignatures' for life. See \cite{olson_oceanographic_2020} for further review of marine biogeochemical cycles as they relate to exoplanet habitability and biosignatures.

\begin{deluxetable}{ll}
\tablenum{2}
\tablecaption{Parameters Varied}
\tablewidth{\textwidth}
\tablehead{
\colhead{Parameter} & \colhead{Value}
\label{tab:params}
}
\startdata
Obliquity & \textit{low:} 0$^\circ$, 15$^\circ$ \\
& \textit{moderate:} 30$^\circ$, 45$^\circ$ \\
& \textit{high:} 60$^\circ$, 75$^\circ$, 90\degree \\
Eccentricity & 0, 0.1, 0.2, 0.3, 0.4
\enddata
\end{deluxetable}

\section{Methods \& Model Description}\label{methods}
Our experiments leverage cGENIE, a 3D marine biogeochemical model, coupled to PlaSim \citep{holden_plasimgenie_2016}. PlaSim is a reduced-complexity 3D atmospheric circulation model, built around the PUMA atmospheric model described by \citet{fraedrich_planet_2005}. PlaSim replaces the 2D energy-moisture balance model (EMBM) typically used in cGENIE simulations, and it provides cGENIE with interactive wind forcing, temperature, humidity, surface pressure, divergence and vorticity. The module also includes fractional cloud cover from relative humidity and precipitation. Time step length is 12 hours for cGENIE and 45 minutes for PlaSim; coupling inputs are averaged over the 16 previous PlaSim time steps to account for this difference \citep{holden_plasimgenie_2016}.

\begin{deluxetable}{cc}
\tablenum{3}
\tablecaption{Instellation Relative to Earth $(S_\oplus)$}
\tablewidth{\textwidth}
\tablehead{
\colhead{Eccentricity} & \colhead{Instellation ($S_\oplus$)}
\label{tab:instellation}
}
\startdata
0.0 & 1.000 \\
0.1 & 1.054 \\
0.2 & 1.082 \\
0.3 & 1.195 \\
0.4 & 1.136 \\
\enddata
\end{deluxetable}

Ocean circulation in cGENIE is handled by C-GOLDSTEIN and GOLDSTEINSEAICE. C-GOLDSTEIN is a reduced physics, frictional geostrophic 3D ocean circulation model that uses advection and diffusion to transport heat, salinity, and biogeochemical tracers \citep{edwards_uncertainties_2005}. Net precipitation minus evaporation (P--E) is represented by a proportionate salinity flux. C-GOLDSTEIN calculates the mixed-layer depth considering the strength of density stratification arising from temperature and salinity gradients and wind stress from PlaSim. GOLDSTEINSEAICE is a dynamic-thermodynamic sea-ice model. Dynamical equations are solved for the percentage of ice cover and the height of ice in each cell. Sea ice growth and decay is governed by heat flux from the atmosphere and ocean. C-GOLDSTEIN, GOLDSTEINSEAICE, and PlaSim are coupled through heat transfer, wind forcing, and the drifting of sea ice along ocean currents \citep{holden_plasimgenie_2016}. 

Chemistry and biology in cGENIE are handled by the ATCHEM and BIOGEM modules for atmospheric chemistry and marine biogeochemistry, respectively \citep{ridgwell_marine_2007}. ATCHEM includes parameterized chemistry involving O$_2$, O$_3$, CH$_4$, and CO$_2$ assuming a Sun-like stellar spectrum \citep{reinhard_oceanic_2020}. ATCHEM passes pCO$_2$ to PlaSim for calculation of surface temperature, but PlaSim does not currently receive pCH$_4$. ATCHEM and BIOGEM are coupled through spatially resolved (2D) sea-air gas exchange that varies with local sea surface saturation state for each species, but ATCHEM assumes a well-mixed atmosphere and homogenizing gaseous species with each time-step. BIOGEM includes photosynthesis, aerobic and anaerobic respiration, methanogenesis \citep{olson_quantifying_2013}, aerobic and anaerobic methanotrophy \citep{olson_limited_2016}, nitrification and sulfide oxidation. Photosynthesis is limited by the availability of both N and P, but N$_2$ fixation occurs when N is scarce relative to P \citep{ridgwell_marine_2007}. Photosynthesis is further constrained by the availability of photosynthetically active radiation (PAR), which is calculated as a fixed fraction of incident radiation. Light is then attenuated in the water column according to an imposed e-folding depth. This calculation implicitly assumes a Sun-like spectrum and that exo-photosynthesis prefers photons of the same wavelength as photosynthesis on Earth.

We run a total of 35 experiments, varying obliquity from 0$^\circ$ to 90$^\circ$ and eccentricity from 0 to 0.4 (Table \ref{tab:params}). Although obliquity simply shapes the spatial distribution of instellation, annual average instellation varies with eccentricity ($e$) by a factor of $(1-e^2)^{-1/2}$ for planets with equivalent semi-major axes (Table \ref{tab:instellation}) \citep{laskar_1993_orbital}. All other parameters---such as day length, surface pressure, and atmospheric pO$_2$ among others---are set to present-day Earth values. We also assume a present-day Earth continental configuration, which allows us to compare the habitability of land vs. marine environments. Finally, we use the Sun's spectrum, which may limit the relevance of our work to more common M-dwarf planets that receive relatively more IR photons and less of the visible photons used by photosynthesis on Earth. This choice is due to current limitations of the photochemistry and photosynthesis codes described above, but it is also physically motivated because HZ planets around M-dwarfs are likely to be tidally locked with near zero obliquity and eccentricity \citep{heller_tidal_2011}. Model runs are spun up for 10,000 years until reaching steady-state with respect to surface temperature and chemical tracers in the ocean. Each experiment requires only 1 CPU and completes in 10 days. Data is output every two weeks, and all data presented herein is averaged over the last decade of each simulation.

\section{Results}\label{results}

\subsection{Extreme surface air temperatures occur locally and seasonally at high obliquity and high eccentricity}

\begin{figure*}
\gridline{\fig{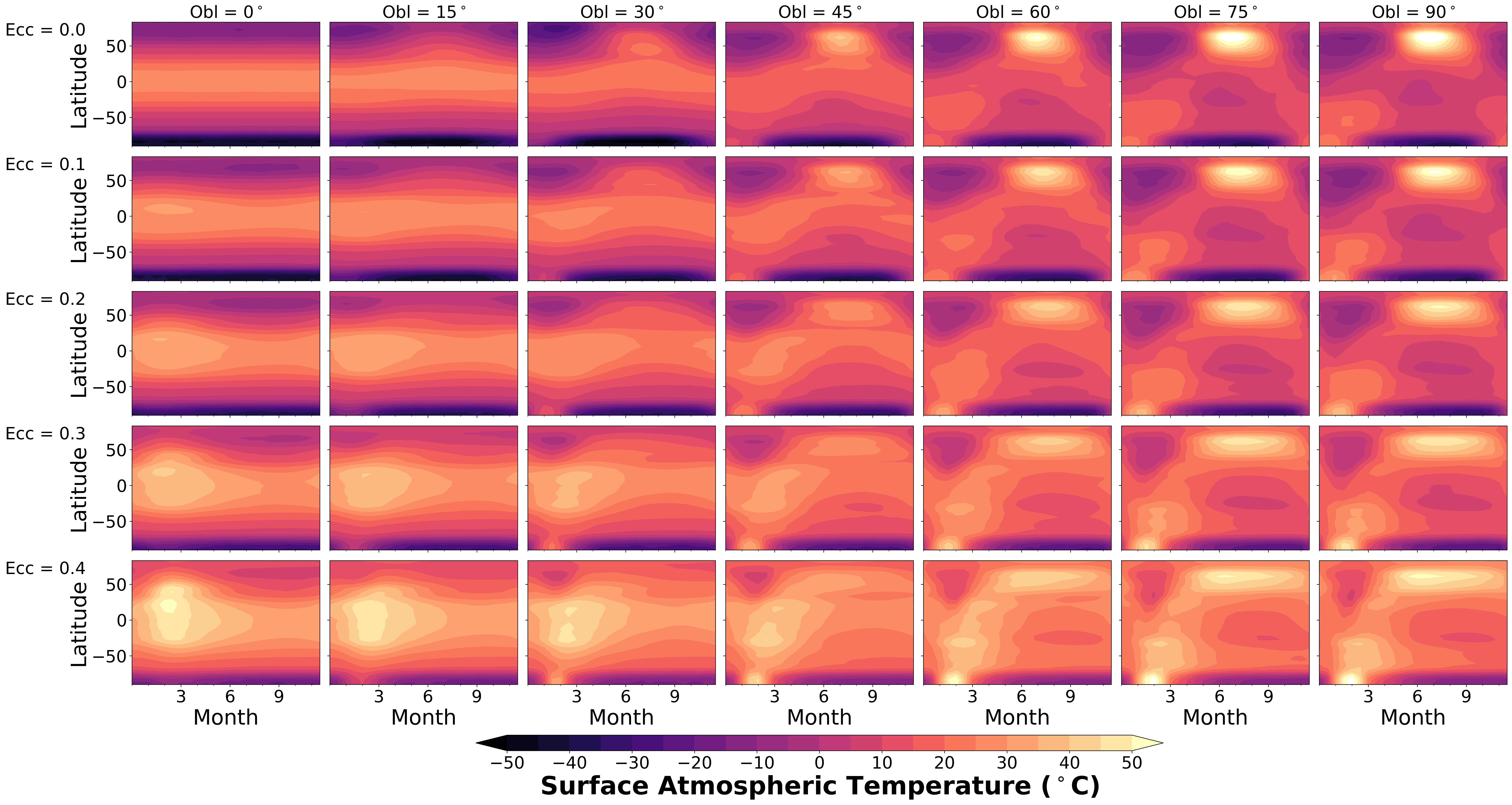}{\textwidth}{}}
\caption{Zonally averaged surface air temperature (SAT) across the seasonal cycle for varied obliquity (columns) and eccentricity (rows). Time advances along the x-axis and latitude is represented on the y-axis for each subpanel. SATs exceed 50$^{\circ}$C during summers for high-obliquity and/or high-eccentricity orbits.}
\label{fig:sat}
\end{figure*}

Surface air temperature (SAT) decreases on annual and global average  by $\sim$5$^\circ$C with increasing obliquity from 0--90$^\circ$ for fixed eccentricity (Figure \ref{fig:sat}). Cooling on global average reflects a $\sim$18$^\circ$C decrease in equatorial SAT, but SAT at the poles actually increases by $\sim$20$^\circ$C with increasing obliquity, leading to reductions in ice cover and planetary albedo as in previous studies. 

Globally and annually averaged SAT increases dramatically with eccentricity for fixed obliquity. SAT increases by $\sim$5$^\circ$C between 0 and 0.2 eccentricity and further increases by $\sim$13$^\circ$C between 0.2 and 0.4 eccentricity on global average. Warming is more pronounced at the poles, which warm by $\sim$20$^\circ$C over this same eccentricity range. This warming on annual average arises along with higher annual-mean instellation on high-eccentricity planets despite asymmetry in season duration (Table \ref{tab:instellation}).

SAT varies only with latitude for planets with both zero obliquity and eccentricity. For all other orbital scenarios, SAT varies both spatially and temporally. Peak SATs surpass 50$^{\circ}$C during the summers for orbits with an eccentricity of 0.4 or an obliquity of $\geq60^{\circ}$, rendering land habitats in the polar regions of planets with high obliquity and the equatorial regions of planets with the combination of high eccentricity and low obliquity habitable only for thermophilic life by Earth standards. Seasonal temperature contrasts of $\geq60 ^\circ$C under these same orbital scenarios may exacerbate this stressor for life on land. However, simulations with both low to moderate obliquity and eccentricity of $\leq$0.3 generate SATs \textless50$^\circ$C year-round at all latitudes, with seasonal temperature variations \textless20$^\circ$C for equatorial and middle latitudes.

\subsection{Sea surface temperatures remain hospitable globally at high obliquity and high eccentricity}

Sea surface temperature (SST) more directly impacts marine life than SAT. Globally and annually averaged SST decreases by $\sim$5$^{\circ}$C from 0$^{\circ}$ to 90$^{\circ}$ obliquity for fixed eccentricity. Globally and annually averaged SST increases by $\sim$4$^{\circ}$C between 0 and 0.2 eccentricity, and it further increases by $\sim$10$^{\circ}$C between 0.2 and 0.4 eccentricity, for fixed obliquity. 

Assessing the habitability of marine environments also requires consideration of how seasonal maxima and minima may affect life locally. Even for the most extreme orbits we consider here, two-week average SST does not exceed 46$^\circ$C at any latitude, suggesting that high obliquity and/or high eccentricity does not preclude marine environments suitable for life. Two-week average SST above 40$^\circ$C occurs only during the summers of simulations with orbits of 0.4 eccentricity and 0$^\circ$--30$^\circ$ obliquity, where it ranges from 43--45$^\circ$C. This muted seasonal maxima relative to SAT arises due to the high heat capacity of water, which makes the ocean more resistant to temperature swings than the atmosphere. The seasonal minima is also muted because SST can not fall below the freezing point of seawater regardless of atmospheric temperature, but we note that permanent sea ice is eliminated for obliquities greater than 45$^\circ$ and eccentricities greater than 0.2 in our simulations. In combination, seasonal SST contrasts are significantly reduced relative to seasonal SAT contrasts. Simulations with low to moderate obliquity and $\leq$0.3 eccentricity display seasonal SST variations of \textless10$^\circ$C. For simulations with both high obliquity and 0.4 eccentricity, seasonal temperature variation in the SH middle and polar latitudes approaches 20$^\circ$C. These seasonal effects are large compared to the Earth, but elimination of polar sea ice nonetheless increases open ocean habitats for life.  

\begin{figure*}
\gridline{\fig{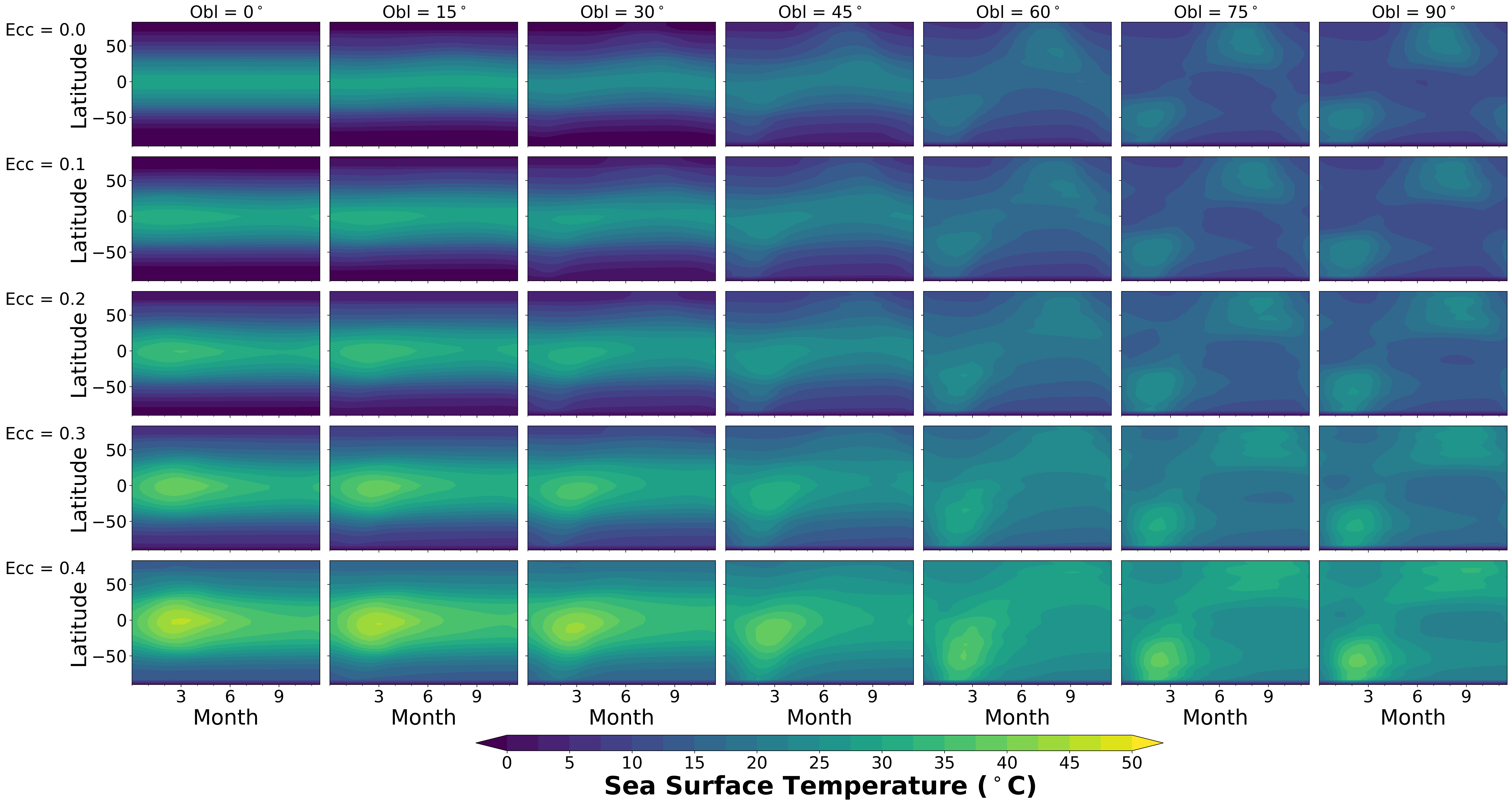}{\textwidth}{}}
\caption{Zonally averaged sea surface temperature (SST) across the seasonal cycle for varied obliquity (columns) and eccentricity (rows). Global average SST increases with increasing eccentricity and decreases slightly with increasing obliquity. Time advances along the x-axis and latitude is represented on the y-axis for each subpanel. Two-week average SST does not exceed 46$^\circ$C for any latitude.} \label{fig:sst}
\end{figure*}

\subsection{Seasonality in the depth of the ocean mixed-layer increases with obliquity and eccentricity}

\begin{figure*}
\gridline{\fig{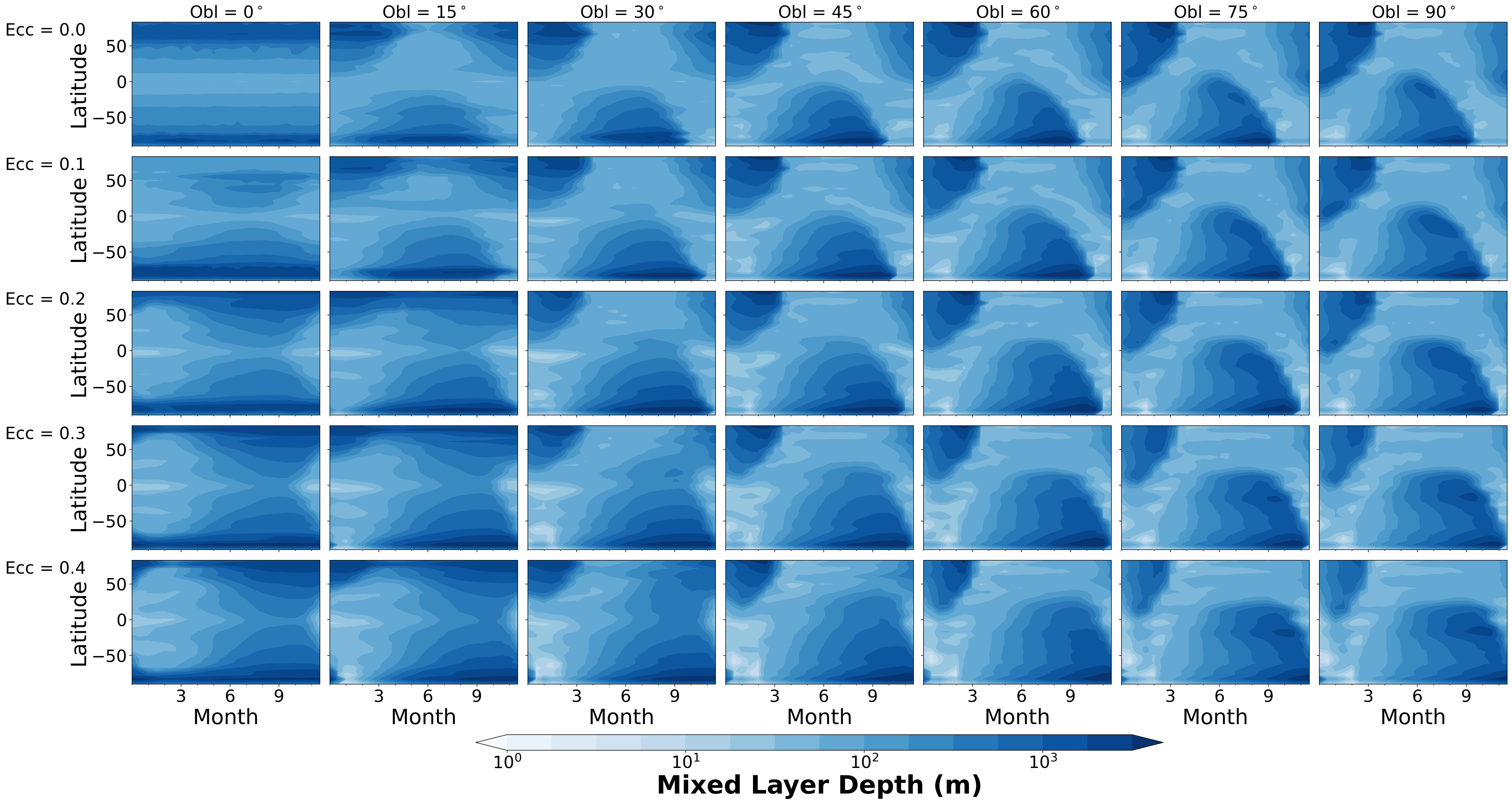}{\textwidth}{}}
\caption{Zonally averaged mixed layer depth across the seasonal cycle for varied obliquity (columns) and eccentricity (rows). Time advances along the x-axis and latitude is represented on the y-axis for each subpanel. Seasonal temperature variations in high-obliquity and, to a lesser extent, high-eccentricity simulations generate extreme fluctuations in mixed layer depth at middle and polar latitudes.}
\label{fig:mld}
\end{figure*}

The depth of the wind-mixed layer varies spatially and temporally with SST (Figure \ref{fig:sst}, \ref{fig:mld}). The mixed layer is shallowest when and where SST is highest, and the mixed layer is deepest when and where SST is lowest. This relationship reflects density stratification of the ocean that opposes vertical mixing, the strength of which varies with latitude and season as the surface of the ocean warms/cools in response to spatially and temporally variable instellation. 

At low obliquities, the mixed layer is shallowest at the equator and deepest at the winter pole. As obliquity increases, the equatorial mixed layer deepens as thermal stratification weakens as the poles begin to receive more stellar energy at the expense of the equator. The depth of the mixed layer also becomes increasingly variable at all latitudes as seasonal temperature contrasts increase. At high obliquities, the depth of the equatorial mixed layer exceeds that of the summer pole (Figure \ref{fig:mld}).

\subsection{Marine biological productivity increases with obliquity and eccentricity}

\begin{figure*}
\gridline{\fig{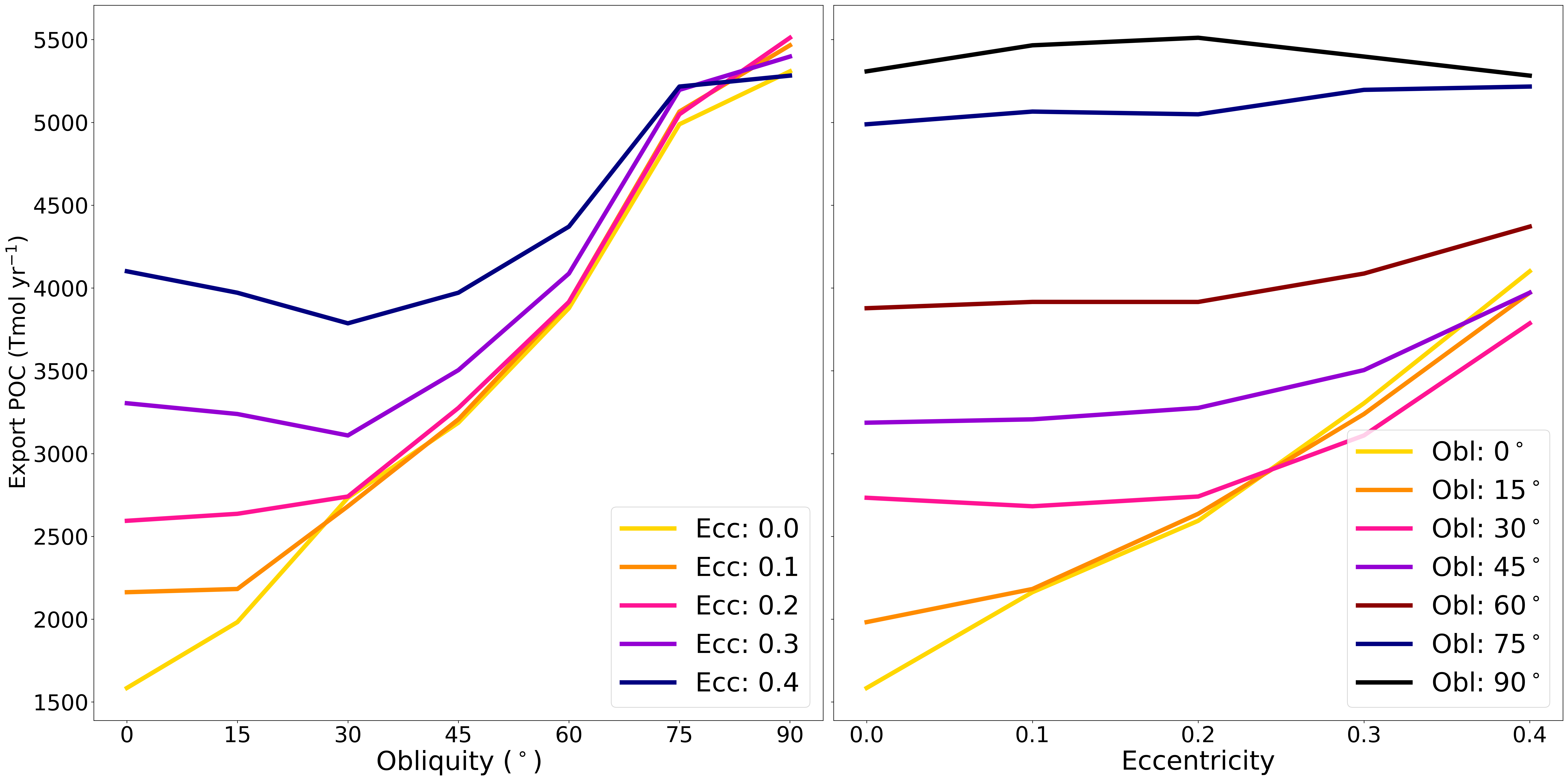}{0.9\textwidth}{}}
\caption{Annual and global average biological export as a function of obliquity and eccentricity. Colored lines represent simulations with identical eccentricities for varied obliquity (left) and identical obliquities for varied eccentricity (right). Export production increases with both increasing obliquity and eccentricity. Eccentricity effects are significant at both low obliquity and eccentricity but drop off as either parameter increases.}
\label{fig:export}
\end{figure*}

Both obliquity and eccentricity influence biological export production. However, obliquity effects are much larger within the parameter space we investigated. At lower obliquity and eccentricity values, increases in both obliquity and eccentricity have large positive effects on export production, but increases in either parameter are less significant at higher values of one or both (Figure \ref{fig:export}). We use our 0$^\circ$ obliquity and 0 eccentricity simulation, which yields 1600 Tmol POL yr$^{-1}$ export production, as a baseline for comparing the biospheric response of worlds experiencing seasons.

For simulations with 0 eccentricity, export production increases nearly linearly with obliquity from 1600 Tmol POC yr$^{-1}$ (baseline) at 0$^\circ$ obliquity to 5200 Tmol POC yr$^{-1}$ (3.3x baseline) at 90$^\circ$ obliquity. Simulations with fixed eccentricities of 0.1 and 0.2 yield small increases in export production over low obliquity ranges from 2200 and 2600 Tmol POC yr$^{-1}$ (1.4 and 1.6x baseline, respectively), converging on 2700 Tmol POC yr$^{-1}$ (1.7x baseline) at 30$^\circ$. At moderate and high obliquities, export production again increases nearly linearly, reaching 5500 Tmol POC yr$^{-1}$ (3.4x baseline) at 90$^\circ$ for these same eccentricities. Simulations with higher eccentricities of 0.3 and 0.4 yield 3300 and 4100 Tmol POC yr$^{-1}$ (2.1x and 2.6x baseline) at 0$^\circ$ obliquity, decrease approximately linearly by 200--300 Tmol POC yr$^{-1}$ over low obliquity ranges, then increase to 5300--5400 Tmol POC yr$^{-1}$ (3.3--3.4x baseline) at 90$^\circ$ (Figure \ref{fig:export}).

For simulations with low obliquity, export production increases nearly linearly with eccentricity, increasing from 1600 and 1900 Tmol POC yr$^{-1}$ (1.0 and 1.2x baseline, respectively) at 0 eccentricity and converge to 4100 Tmol POC yr$^{-1}$ (2.6x baseline) at 0.4 eccentricity. 

For moderate and high obliquities, export production is initially higher at 0 eccentricity and is less sensitive to eccentricities $\leq$0.2, changing by $\leq$4\%. In simulations with 30--60$^\circ$ obliquity, export production then increases nearly linearly from 0.2 up to 0.4 eccentricity, yielding 3800--4300 Tmol POL yr$^{-1}$ (2.4--2.7x baseline) at 0.4 eccentricity. Meanwhile, export production in higher obliquity simulations with 75$^\circ$ and 90$^\circ$ obliquity remains relatively insensitive to eccentricity. Export increases by only 100 Tmol POL yr$^{-1}$ in our 75$^\circ$ simulation, while export decreases by 400 Tmol POL yr$^{-1}$ in our 90$^\circ$ simulation, with both scenarios converging to $\sim$5200 Tmol POL yr$^{-1}$ at 0.4 eccentricity (Figure \ref{fig:export}).

\begin{figure*}
\gridline{\fig{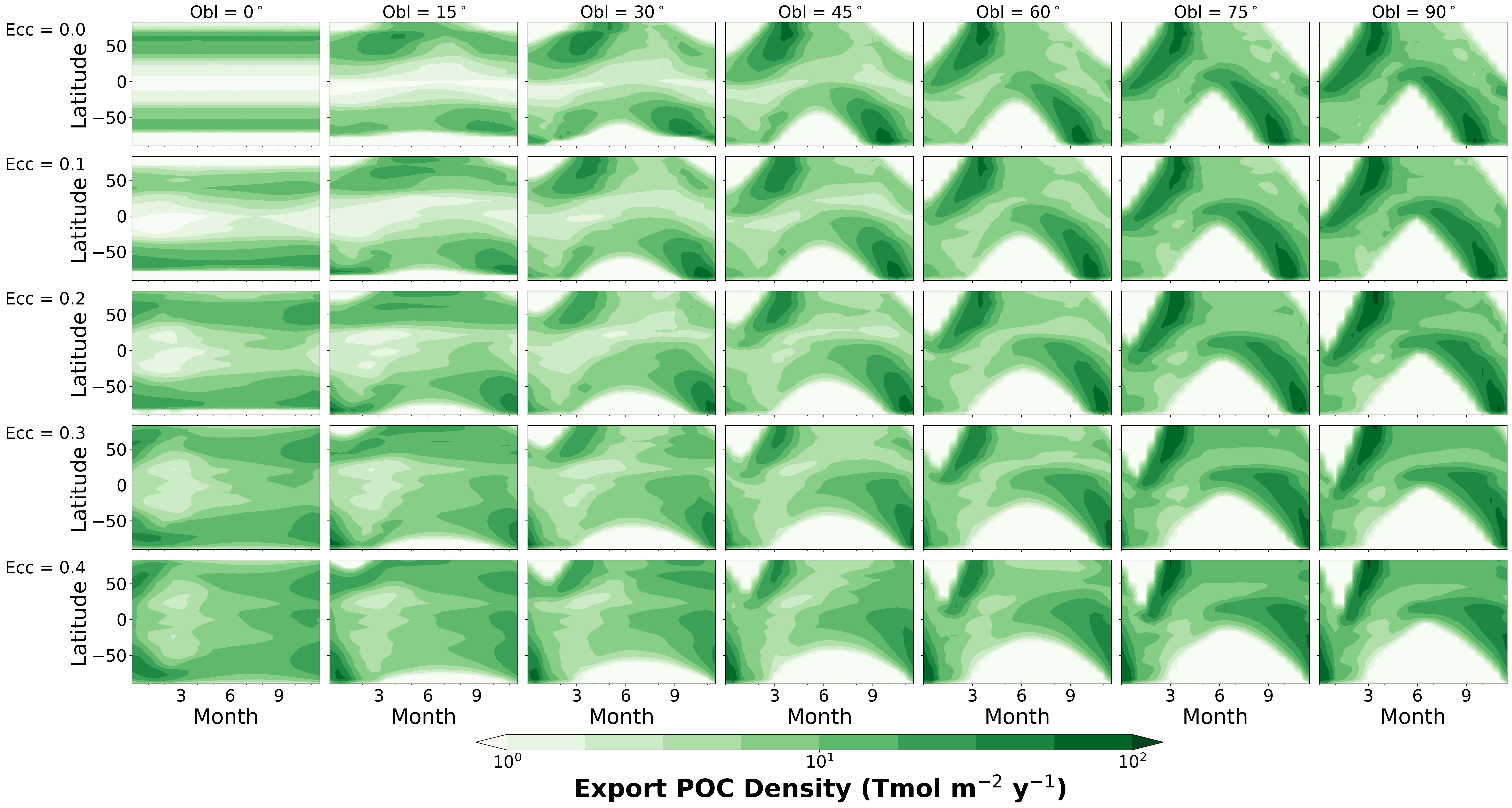}{\textwidth}{}}
\caption{Zonally averaged export density across the seasonal cycle for varied obliquity (columns) and eccentricity (rows). Time advances along the x-axis and latitude is represented on the y-axis for each subpanel. Nearly all biological export occurs in latitudes where the depth of the mixed layer varies seasonally, just after the transition from a deep to a shallow mixed layer.}
\label{fig:export_hov}
\end{figure*}

The distribution of export production is closely related to mixed-layer depth, both spatially and temporally. In every simulation, nearly all export production occurs at latitudes with high mixed-layer depth seasonality, during and immediately following the transition from deep to shallow mixed layer as thermal stratification develops with surface warming (Figures \ref{fig:mld}, \ref{fig:export_hov}), a phenomenon referred to as `spring blooms' on Earth. This relationship between productivity and mixed-layer depth arises due to seasonal differences in nutrient availability as ocean stratification varies \citep{barnett_moderately_2022}. 

\section{Discussion}\label{discussion}

\subsection{Support and extension of previous work}
\citet{barnett_moderately_2022} showed that annual biospheric productivity increases with obliquity in cGENIE simulations, and they argued that moderately high obliquity planets may thus be particularly attractive candidates for exoplanet life detection. However, limitations to their modeling methodology precluded them from investigating obliquities higher than 45$^{\circ}$ and it was unclear whether the trend of increasing productivity with obliquity would continue to more extreme orbits. We have improved upon their initial study by coupling the PlaSim atmospheric GCM to cGENIE to allow for seasonal reversal of surface winds, and we demonstrate that biospheric productivity increases with obliquity up to 90$^{\circ}$ (Figure \ref{fig:export}), suggesting that high-obliquity planets may be superhabitable \citep{heller_superhabitable_2014}. We also show that seasonality arising from high-eccentricity orbits also favorably influences biospheric productivity (Figure \ref{fig:export}). Perhaps most surprising, the combination of high obliquity and high eccentricity supports marine biospheres that are more productive than present-day Earth in our model (Figure \ref{fig:export})!

As concluded in previous work, ocean life responds favorably to `extreme' orbits in our simulations due to changes in the distribution and recycling of bioessential nutrients in the oceans \citep{olson_oceanographic_2020, barnett_moderately_2022}. Enhanced nutrient availability in our simulations is the consequence of seasonal breakdown of density stratification within the ocean (Figure \ref{fig:mld}), allowing nutrient-rich deep waters to mix with nutrient-depleted surface waters and stimulating photosynthetic activity. 

\subsection{Implications for exoplanet life detection} 

Higher productivity on high-obliquity and high-eccentricity planets may favorably influence the `detectability' of remote biosignatures. Higher productivity results in greater production of photosynthetic O$_2$, a frequently discussed biosignature that signals the presence of life on Earth \citep{meadows_reflections_2017, meadows_exoplanet_2018, schwieterman_exoplanet_2018}. Moreover, higher rates of photosynthesis ultimately increase the potential rates of other metabolisms in the ocean interior and/or marine sediments---and thus the production of other metabolic waste products that may also serve as gaseous biosignatures. This relationship arises because photosynthesis translates stellar energy to chemical energy that is then propagated through the biosphere in the form of chemical disequilibrium \citep{krissansen-totton_disequilibrium_2018}. This energy flux on Earth is orders-of-magnitude greater than geological energy inputs, with the consequence that Earth's biosphere is ultimately solar-powered despite a diversity of non-photosynthetic metabolisms. Photosynthetic rates thus throttle the production of most biogenic gases that may serve as remote biosignatures, including those not directly involved in photosynthesis (e.g., CH$_4$, N$_2$O, H$_2$S).

Biosignature gases may be produced in large quantities without influencing planetary spectra, potentially resulting in a `false negative' for life \citep{reinhard_false_2017}. As an example, microbial sulfate reduction produces large fluxes of H$_2$S in marine sediments on present-day Earth \citep{canfield_sulfate_1991}, but the resulting H$_2$S forms pyrite (FeS$_2$) within sediments or is re-oxidized before it can reach the atmosphere and influence the spectral appearance of our planet. Likewise, the Black Sea---an analog environment for Earth's ancient oceans---supports high rates of methanogenesis, but it is ultimately a negligible source of CH$_4$ to the atmosphere due to oxidation of CH$_4$ within the water column \citep{schmale_response_2011}. Alien observers expecting to find H$_2$S or CH$_4$ on inhabited planets may thus be mislead by their low-abundances in our atmosphere. 

Exoplanets with large seasonal cycles may be less vulnerable to these types of false negatives. In addition to replenishing nutrients, seasonal breakdown of the thermal stratification that inhibits vertical mixing in the ocean on high-obliquity and high-eccentricity worlds may increase the fraction of biosignature gases produced in the ocean interior and marine sediments that ultimately make it to the surface and accumulate in the atmosphere where they may affect planetary spectra. The combination of greater biosignature production and enhanced communication between the ocean interior and the atmosphere may thus increase the detectability of life on worlds experiencing large seasonal cycles. 

In addition to increasing the atmospheric accumulation of biosignature gases, seasonality may also serve as an independent beacon for life \citep{olson_atmospheric_2018, mettler_earth_2022}. The composition of Earth's atmosphere oscillates seasonally, reflecting seasonally variable gas fluxes as life responds to its changing environment. Most famously, CO$_2$ levels rise and fall at Mauna Loa as the balance between CO$_2$ fixation into biomass by photosynthesis and CO$_2$ release by respiration shifts with the seasons \citep{keeling_atmospheric_1976}. Similar fluctuations on other worlds may provide a temporal biosignature that reveals the presence of exoplanet life. 

These seasonal changes are small on present-day Earth but may be exaggerated on worlds with larger seasonal cycles. However, it remains to be seen whether obliquity- or eccentricity-driven seasonality in atmospheric composition will be more detectable. Our simulations suggest that obliquity is more likely to result in large-magnitude biogenic seasonality compared to eccentricity, but characterizing seasonality on these worlds is complicated by the details of viewing geometry \citep{olson_atmospheric_2018, mettler_earth_2022}. In some scenarios, views that blend opposing seasons between hemispheres may mask the biogenic signal in disk average. Global seasons on planets with eccentric orbits may thus be easier to characterize than hemispheric seasons resulting from obliquity for many viewing geometries, but our results imply that detectable seasonality may require very high eccentricities compared to Earth's. 

A recent study by \cite{schulze-makuch_search_2020} generated a list of two dozen potentially superhabitable exoplanets and exoplanet candidates from the Kepler catalogue, taking into account various planet and host star properties previously known to favorably influence habitability \cite{heller_superhabitable_2014}. Our results suggest that high-obliquity and/or high-eccentricity worlds should also be considered among the most favorable targets for life detection. By extension, observationally constraining obliquity and eccentricity will be critical for characterizing habitability and for the interpretation of biosignatures. While eccentricity can be derived from radial velocity measurements and low-resolution light curves prior to atmospheric characterization measurements, constraints on planetary obliquity require more detailed and frequent measurements \citep{kipping_transiting_2008, van_eylen_eccentricity_2015, kane_obliquity_2017}. Previous studies have derived planetary obliquity from full thermal phase curves of large hot Jupiters \citep{adams_signatures_2019}. Thermal emission may also reveal the obliquities of directly imaged young jovian planets \citep{cowan_light_2013}, but resolving the obliquities of mature planets with less internal heat is complicated by several additional factors \citep{gaidos_seasonality_2004}. The obliquities of Earth-sized rocky exoplanets are inaccessible with current instrumentation \citep{kane_obliquity_2017}, but full-orbit reflected light curves obtained with future instrumentation may one day constrain the obliquities of potentially habitable worlds \citep{kawahara_mapping_2011}. Knowledge of planetary obliquity could provide important context for interpreting atmospheric observations relating to both planetary habitability and biosignatures. This context may provide an opportunity to evaluate exoplanets for future observations and prioritize those that have the highest potential for the detection of biosignatures.
Future work and/or instrument design should consider constraining the obliquities of rocky planets in the HZ of Sun-like stars a priority. 

\subsection{Caveats and opportunities for future work}
A potential caveat is that the same compositional seasonality that may signal the presence of exoplanet life may also represent a challenge for certain types of life. For example, seasonally variable ocean chemistry may threaten the development of animal-grade complexity if such life requires stable oxygenation of its environment \citep{catling_why_2005,reinhard_earths_2016}. Indeed, ocean deoxygenation is associated with several of the `big 5' mass extinctions on Earth, highlighting the potential impact of varying oxygen levels on marine animals \citep{meyer_oceanic_2008}. However, if evolution via natural selection favors individuals tolerant of seasonal changes in ocean composition on high-obliquity and/or high-eccentricity worlds, then marine animals may be more resilient against the types of perturbations associated with mass extinctions on Earth. Future work should investigate how seasonality in the composition of the ocean-atmosphere system may specifically affect animals, which are not represented in our simulations. 

The long-term effect of obliquity on planetary habitability is also an area of ongoing research. Previous studies have shown that increasing planetary obliquity can extend the outer edge of the HZ through enhanced heating at the poles due to increased seasonality, thereby increasing the orbital distance an exoplanet with high obliquity can occupy around its host star and remain habitable \citep{williams_habitable_1997, spiegel_habitable_2009, dressing_habitable_2010, armstrong_effects_2014, linsenmeier_climate_2015, wang_effects_2016, kilic_stable_2018, nowajewski_atmospheric_2018, guendelman_atmospheric_2019, kang_mechanisms_2019, colose_enhanced_2019, palubski_habitability_2020, komacek_constraining_2021}.
However, a GCM study by \citet{kang_wetter_2019} showed that planets inside the HZ can experience higher stratospheric water vapor concentrations due to the effects of high obliquity. 
Higher concentrations of stratospheric water vapor could impact the long-term habitability of high-obliquity planets through enhanced water loss \citep{kasting_runaway_1988, meadows_reflections_2017}. 
While high-obliquity planets are attractive candidates for future observation because of their potential to increase biological productivity, enhanced water loss could create false-positive biosignature detections from the subsequent accumulation of abiotic oxygen or potentially threaten long-term habitability \citep{meadows_exoplanet_2018}. Future work is needed to constrain the effect of obliquity on planetary water loss to discern the relative advantage of higher obliquity on overall planetary habitability.

\section{Conclusions}\label{conclusions}
We used cGENIE-PlaSim, a 3D marine biogeochemical model coupled to an atmospheric GCM, to explore the habitability of high-obliquity and high-eccentricity planets. We considered a large range of orbital scenarios including simulations with 0--90$^\circ$ obliquity and/or 0--0.4 eccentricity. Sea surface temperatures consistently remained within the bounds considered hospitable for mesophilic life on Earth at all latitudes for the entire year in nearly all of our model scenarios, despite extreme surface air temperatures in some simulations. We therefore conclude that marine life could thrive year-round on high-obliquity and/or high-eccentricity worlds. 

In fact, our simulations suggest that seasonal variations driven by obliquity and/or eccentricity can be beneficial for life. Biospheric productivity on a high-obliquity planet typically exceeds that on an otherwise identical low-obliquity planet in our model. The same holds true for high-eccentricity versus low-eccentricity worlds, although the observed increase in productivity with increasing eccentricity is comparatively subdued. In both cases, the mechanism by which productivity increases is enhanced nutrient recycling via seasonal breakdown of thermal stratification within the ocean, consistent with previous studies \citep{olson_oceanographic_2020,barnett_moderately_2022}. Higher rates of photosynthesis introduce more chemical energy to the global biosphere, benefiting many metabolisms in addition to photosynthesis. High-obliquity and high-eccentricity planets may therefore be `superhabitable' worlds capable of supporting larger and possibly more diverse biospheres than Earth. 

Our results also suggest that high obliquity and high eccentricity can favorably influence biosignatures in addition to habitability. Higher biospheric productivity may ultimately increase the annual production of biosignature gases such as O$_2$ and CH$_4$ on high-obliquity and high-eccentricity planets, minimizing the risk of false negative non-detections. At the same time, seasonal oscillations in biological activity on these planets may manifest as remotely detectable oscillations in atmospheric composition. This time-variability provides an independent temporal biosignature unique to planets experiencing seasons that may mitigate against false-positives and false-negatives \citep{olson_atmospheric_2018, schwieterman_exoplanet_2018}. 

Our study suggests that high-obliquity and high-eccentricity planets may be appealing targets for exoplanet life detection. However, there are still some open questions that require urgent attention. For instance, the extent to which seasonality affects the development of complex life remains to be explored. Long-term water loss and its implications for both habitability and biosignature false positives on high-obliquity planets may also be a concern. Moving forward, observationally constraining planetary obliquity should also be a priority. With these caveats, we argue that high-obliquity and high-eccentricity planets may be among our most promising targets for the search for exoplanet life.

\begin{acknowledgments}
SLO acknowledges support from the NASA Exobiology, Habitable Worlds, and ICAR programs under grants 80NSSC20K1437, 80NSSC20K1409, and 80NSSC21K0594, respectively. We also thank Christopher Colose and René Heller for providing helpful comments to improve this manuscript. This project benefited from participation in the NASA NExSS and NASA NOW Research Coordination Networks. 
\end{acknowledgments}

\bibliography{orbits}{}
\bibliographystyle{aasjournal}

\end{document}